\def\1{\'{\i}}
\def\2{\~n}

\def\matriz#1#2{\left( \begin{array}{#1} #2 \end{array}\right) }
\def\R{\mathbb R}

\def\a{\alpha}
\def\b{\beta}
\def\g{\gamma}

\def\s{\sigma}
\def\l{\lambda}
\def\n{\nu}

\def\om{\omega}
\def\vp{\varphi}
\def\r{\rho}

\def\th{\theta}
\def\D{\Delta}
\def\L{\Lambda}

\def\pd#1#2{\frac{\partial#1}{\partial#2}}

\def\raiz{\sqrt{2 p_1}}

\newtheorem{theorem}{Theorem}
\newtheorem{proposition}{Proposition}

\documentclass[10pt]{iopart}
\usepackage{iopams}  
\usepackage{times}
\usepackage{cite}
\begin{document}

\title[]{Poisson structures for reduced non-holonomic systems}
\author{Arturo Ramos
\footnote[1]{To whom correspondence should be addressed (aramos@math.unipd.it)}
}

\address{
Dipartimento di Matematica Pura ed Applicata,
Universit\`a degli Studi di Padova, \\ Via G. Belzoni 7, I-35131 Padova, Italy}

\address{\ E-mail: aramos@math.unipd.it}

\begin{abstract}
Borisov, Mamaev and Kilin have recently found certain 
Poisson structures with respect to which the reduced 
and rescaled systems of certain non-holonomic problems,
involving rolling bodies without slipping, 
become Hamiltonian, the Hamiltonian function 
being the reduced energy. We study further the algebraic 
origin of these Poisson structures, showing that they 
are of rank two and therefore the mentioned rescaling is
not necessary. We show that they are determined, up to a
non-vanishing factor function, by the existence of 
a system of first-order differential equations providing
two integrals of motion.
We generalize the form  of that Poisson structures 
and extend their domain of definition. 
We apply the theory to the rolling disk, 
the Routh's sphere, the ball rolling 
on a surface of revolution, and its special case
of a ball rolling inside a cylinder.
\end{abstract}

\pacs{02.40.k, 03.04.t}

\ams{70G45, 70E18, 70F25}

\submitto{\JPA}


\section{Introduction}

In recent years there has been an increasing interest in the geometric 
treatment of non-holonomic mechanical systems, see, e.g., 
\cite{Bat02,BatCus99,BatSni92b,BloKriMarsMur96,CanLeoMarrMar98,CarFav96,
CusKemSniBat95,Marle95,Marle96b,Marle98a,Sni98,Sni01,Sni02}. 
In particular, it has been recognised that the Hamiltonian 
formulation of such systems can be stated in terms of 
an \emph{almost-Poisson} bracket, that is,
a biderivation of functions of phase space, antisymmetric in its 
arguments but which does not necessarily fulfil the 
Jacobi identity (see, e.g., 
\cite{Bat98b,CanLeoMar99,SchMas94}).
Therefore, for researchers in this field, 
it seems to be usual the conceptual association of the Hamiltonian formulation
of non-holonomic mechanical systems with almost-Poisson structures. 

On the other hand, there exist non-holonomic systems which,
after certain reductions are performed, admit a Hamiltonian 
formulation after a \lq\lq rescaling of time\rq\rq\ is carried out,
by means of rescaling factors (sometimes called \emph{invariant measures})
of the reduced vector field of the system. This is the case
for the so-called LR systems, which are systems 
formulated on compact Lie groups endowed with a left-invariant 
metric and right-invariant non-holonomic constraints. 
After a rescaling of time, their corresponding reduced systems 
become integrable Hamiltonian systems describing geodesic 
flows on unit spheres \cite{FedJovb}. 
In \cite{CanCorLeoMar02}, a necessary and sufficient condition 
for the existence of an invariant measure for the reduced dynamics 
of generalized Chaplygin systems of mechanical type is given. 
Another recent work on this line is \cite{ZenBlo03}.
For a classic treatment of the theory of Chaplygin's reducing
multiplier, see Section III-12 of \cite{NeiFuf72}.
Thus, it could be conceptually associated as well the existence 
of specific rescaling factors for these reduced systems with the
possibility of formulating them in a Hamiltonian way. 

In addition, Borisov, Mamaev and Kilin \cite{BorMam02c,BorMamKil02}
have recently found a Poisson structure for each studied case of
reduced non-holonomic systems, such that the reduced system becomes
Hamiltonian, with respect to such a structure, after a rescaling,
the Hamiltonian function being the reduced energy. The examples
treated by them are classical in the literature, consisting mainly of 
rolling bodies without slipping, namely  
a rigid body of revolution rolling on a plane, in particular
the \emph{Routh's sphere} (see Section~\ref{sect_Routh_sphere}),
the rolling disk (to be treated in Section~\ref{sect_rolling_disk}),
the motion of a homogeneous ball on a surface of revolution 
(called sometimes \emph{Routh's problem}, see, e.g., \cite{Zen95,ZenBlo03}), 
and other cases. 
There is a strong emphasis in these references in the 
sense that the Poisson structure for each case can be found 
\emph{after a rescaling of time} of the reduced vector field.

Our primary motivation for this work was to understand the origin 
of the two integrals of motion appearing in the mentioned problem of a 
ball rolling without slipping inside a surface of revolution, 
which are not given, in general, in an explicit form but being 
related to the solutions of a system of first order non-autonomous 
differential equations \cite{Rou60,Herm95b,Zen95}. This also happens 
in the other mentioned cases.
The results of \cite{BorMam02c,BorMamKil02} suggest that such 
systems can be interpreted as the equations providing a set of 
functionally independent Casimir functions of the Poisson 
structure they find for each specific case. 
Therefore, it seemed to be worth investigating further such
Poisson structures, in particular to clarify their domain of 
definition and basic properties. Let us note that 
another recent approach, devoted to the study of Poisson 
structures which can be associated to never vanishing vector 
fields on manifolds of arbitrary dimension $d\geq 2$, 
with \emph{fibrating periodic flows}, is given in \cite{FasGiaSan04}. 

It follows that the previously mentioned Poisson structures 
have a rather peculiar form. In particular, the associated characteristic 
distributions have rank two in the open sets of the reduced 
spaces considered in \cite{BorMam02c,BorMamKil02}.
This property implies that such Poisson structures, when multiplied 
by a never vanishing function, are again Poisson structures 
of the same type. The immediate consequence is that the 
above mentioned reduced non-holonomic systems are already Hamiltonian
with respect to one of these Poisson structures without any need of rescaling. 

Other interesting result is that, in the cases studied, 
the Poisson structures obtained can be extended {}from their
original domains of definition, namely (open sets of) 
semialgebraic subvarieties of $\R^5$, to an open set of 
the ambient space. Such extended Poisson structures 
become zero only at the so-called \emph{singular equilibria} 
of the reduced systems. Moreover, the existence of 
these (extended) Poison structures, from an algebraic point of 
view, is only caused by the existence of integrals of motion 
of the reduced vector field related to the solutions 
of the mentioned systems of first order differential equations. 

This paper is organized as follows. In Section~\ref{pois_str_rank2}
we briefly review some notions of Poisson geometry and in particular,
of Poisson structures of rank two. In Section~\ref{pois_str_rank2}
we show the explicit expressions of certain bivectors
in $\R^4$ and $\R^5$, determined up to a non-vanishing factor function,
by choosing the 1-forms in their kernels to have a specific form,
and we prove that they are in fact Poisson bivectors of rank two.
Section~\ref{sect_of_examples} is devoted to show the application
of the previous results in specific examples, namely, the rolling disk, 
the Routh's sphere, and the ball rolling on a surface of revolution.
We will use the formulation of \cite{CusHerKem96}, \cite{Cus98} and \cite{Herm95b}, 
respectively, of these problems, rather than that of \cite{BorMam02c,BorMamKil02}. 
However, we point out the equivalence of both treatments in the last case.
We also treat the special case of a ball rolling inside a cylinder.
Finally, we end with some conclusions and an outlook for further research.

\section{On Poisson structures of rank two\label{pois_str_rank2}} 

For the sake of completeness and in order to fix some notations,
we will recall some well-known notions on Poisson manifolds, 
and in particular, we will focus on Poisson structures of rank two. 
For more details see, e.g., \cite{LibMarle87}. 

Given a differentiable manifold $M$, a \emph{Poisson structure}
on $M$ is defined by an antisymmetric bilinear map $\{\cdot,\cdot\}$ 
which is a derivation on both of its arguments, satisfying moreover 
the Jacobi identity. A manifold $M$ endowed with a Poisson structure 
is called a \emph{Poisson manifold}. 

Thus, it is possible to associate to each function $f$ a 
unique vector field $X_f$ such that, for any 
other function $g$, we have $X_f g=\{f,g\}$. The vector field
$X_f$ is called \emph{Hamiltonian vector field} associated to
the \emph{Hamiltonian function} $f$. This association defines
an homomorphism of the Lie algebra $(C^\infty(M),\{\cdot,\cdot\})$ 
onto the Lie algebra of vector fields in $M$. A \emph{Casimir function}
or \emph{Casimir} for short, is a function $c$ such that $X_c=0$.
The Poisson structure is called \emph{non degenerate} if only
the constant functions are Casimir functions.

Moreover, on every Poisson manifold, there exists a unique 
twice contravariant anti\-sym\-me\-tric tensor field 
(called \emph{bivector field} for short) $\L$ such that
$\{f,g\}=\L(df,dg)$ for every pair of functions $(f,g)$.
This tensor field is called the \emph{Poisson tensor} of the
structure, and the manifold $M$, endowed with its Poisson structure,
will be denoted $(M,\L)$. The existence of such a tensor field is due
only to the antisymmetry and derivation properties of the Poisson 
bracket. The fulfillment of the Jacobi identity for the Poisson
bracket is equivalent \cite{Lich77} to the vanishing of the 
\emph{Schouten--Nijenhuis bracket} 
of $\L$ with itself, $[\L,\L]=0$. The Schouten--Nijenhuis
bracket \cite{Scho54,Nij55} is the unique extension of the 
Lie bracket of vector fields to the exterior algebra 
of multivector fields. Some of its 
properties are   
\begin{eqnarray}
&& [P,Q]=-(-1)^{(p-1)(q-1)}[Q,P] \nonumber\\
&& [P,Q\wedge R]=[P,Q]\wedge R+(-1)^{(p-1)q} Q\wedge[P,R] \label{props_Scho}\\
&& [P\wedge R,Q]=P\wedge[R,Q]+(-1)^{(q-1)r} [P,Q]\wedge R  \nonumber
\end{eqnarray}
where $P,Q,R$ are completely antisymmetric contravariant tensors of
degree $p,q,r$, res\-pec\-ti\-ve\-ly. For more details and properties
on the Schouten--Nijenhuis bracket see, e.g., \cite{CarIboMarPer94,Marle97,Nij55,Scho54} 
and references therein.

Take a local chart of $M$, with domain $U$ and associated
local coordinates $(x_1,\dots,x_n)$, where $n={\rm dim}\,M$.
We will denote by $\L_{ij}$, ($1\leq i,j\leq n$) the components
of the Poisson tensor $\L$ in the previous chart. 
The expression of the Poisson bracket of the restriction 
of the two functions $f,g$ to $U$, also denoted by $f,g$, reads
\begin{eqnarray}
\{f,g\}=\L_{ij}\pd{f}{x_i}\pd{f}{x_j}\,,
\nonumber
\end{eqnarray}
where summation in the repeated indices is understood. In particular we
have $\{x_i,x_j\}=\L_{ij}$. The Poisson tensor admits the local expression
\begin{equation}
\L=\sum_{i<j}^n\L_{ij}\pd{}{x_i}\wedge\pd{}{x_j}
\label{Poiss_biv_loc}
\end{equation}
in these coordinates. 

Given a Poisson manifold $(M,\L)$, it can be defined the 
fibered morphism $\L^\sharp: T^*M\rightarrow TM$ such that 
for any pair of 1-forms $\a$, $\b$, 
$\langle\L^\sharp(\a),\b\rangle=\L(\a,\b)$. 
The image of the morphism $\L^\sharp$, $C=\L^\sharp(T^*M)$, 
is called \emph{the characteristic distribution} of the Poisson structure,
and the \emph{characteristic space} on $x\in M$ is the vectorial 
subspace $C_x=\L^\sharp_x(T^*_xM)$ of $T_xM$. The \emph{rank} of the 
structure on the point $x$ is the rank of $\L_x^\sharp$, i.e., the 
dimension of $C_x$. Note that the annihilator of the characteristic distribution,
i.e., $C^0=\{\b\in\L^1(M)\ |\ \L(\b,\a)=0\,, \forall\a\in\L^1(M)\}$,
is ${\rm ker}\,\L^\sharp$, and we have
${\rm rank}\,\L_x^\sharp+{\rm dim}\,{\rm ker}\,\L_x^\sharp=n$, 
for all $x\in M$. In general, the rank of the structure
varies with $x$ and thus $C$ is not in general a subbundle of $TM$.

Consider now a Poisson manifold $(M,\L)$, ${\rm dim}\,M=n$, such 
that in the domain of a local chart $(U,\phi)$ the structure has constant 
rank equal to two. The Theorem 11.5 of Chapter III in \cite{LibMarle87} 
(or Corollary 2.3. in \cite{Wei83})
assures us that the associated local coordinates, denoted $(x,y,z_1,\dots,z_{n-2})$,
can be chosen such that for $1\leq k\,,l\leq n-2$,
\begin{equation}
\{y,x\}=1\,,\quad\{x,z_k\}=0\,,\quad\{y,z_k\}=0\,,\quad\{z_k,z_l\}=0\,.
\label{pois_prop}
\end{equation}

We are now in a position to prove a simple result, but important for 
our purposes here: 
\begin{proposition}
Let $(M,\L)$ be a Poisson manifold of (locally) constant rank equal to two.
Then, for each never vanishing smooth function $a\in C^\infty(M)$, 
$(M,a\L)$ is a Poisson manifold of (locally) constant rank equal to two, 
with the same characteristic distribution.  
\label{prop_conf}
\end{proposition}

{\noindent {\it Proof}} 

We have to prove that the Schouten--Nijenhuis bracket $[a\L,a\L]$ vanishes,
the other needed properties being obvious. {}From the paragraph 18.8 of 
Chapter V of \cite{LibMarle87}, we have that
$$
[a\L,a\L]=2 a\, \L^\sharp(da)\wedge \L\,. 
$$
It suffices to compute the previous expression on 
a coordinate neighbourhood like that described in the previous 
paragraph, with respect to the Poisson tensor $\L$ \cite{Marle97}. We have
$$
[a\L,a\L](dx,dy,dz_k)
=2 a (\L^\sharp(da)\wedge \L)(dx,dy,dz_k)=0\,,\quad 1\leq k\leq n-2\,,
$$
because $z_k$ are Casimir functions of $\L$, and $dz_k$ enter at 
least once as argument of $\L$ in all terms of the previous expression.
For other possible arguments the expression vanishes by the same reason.

\medskip 
{\noindent {\bf Example 1}} 
Let $M$ be a $n$-dimensional manifold and $X$, $Y$ two 
vector fields such that for all $x\in M$, the Lie bracket 
$[X,Y]_x$ belongs to the subspace of $T_xM$ generated
by $X_x$ and $Y_x$. Then, $\L=X\wedge Y$ is a Poisson 
tensor of rank two except where $X$ and $Y$ are 
linearly dependent. This is easily seen by deducing 
{}from the properties of the Schouten--Nijenhuis 
bracket (\ref{props_Scho}) the relation 
$[X\wedge Y,X\wedge Y]=2 X\wedge Y\wedge [X,Y]$, 
see also \cite{Bat98b,CarIboMarPer94}.

\medskip 
{\noindent {\bf Remark}} Note that it is essential 
in Proposition~\ref{prop_conf} the assumption that the initial 
bivector is Poisson, which assures the existence of 
local coordinates satisfying (\ref{pois_prop}). 
The existence of a bivector whose rank is always two is not enough
to conclude that it is a Poisson bivector. A simple counter-example 
is the following. Take $M=\R^3$, with coordinates $(x,y,z)$.
Let $X$, $Y$ be vector fields given by 
$$ 
X=\pd{}{x}-y\pd{}{z}\,,\quad Y=\pd{}{y}+x\pd{}{z}\,.
$$
Then, $\L=X\wedge Y=\pd{}{x}\wedge\pd{}{y}$ is an 
everywhere rank two bivector but is not Poisson, 
since $[X,Y]=2\pd{}{z}$ and 
$$
[\L,\L]=[X\wedge Y,X\wedge Y]
=4 \pd{}{x}\wedge\pd{}{y}\wedge\pd{}{z}\,.
$$
The vector fields $X$, $Y$ and $[X,Y]$ in this example 
close on a Lie algebra isomorphic to the Heisenberg--Weyl 
Lie algebra ${\mathfrak h}(3)$, see, e.g., \cite{CarRam03}.

\section{Some Poisson structures of rank 
two in $\R^4$ and $\R^5$\label{Pois_struc_R4_R5}}

We will construct in this Section some Poisson structures of rank
two in $\R^4$ and $\R^5$ by imposing that the kernel of the
corresponding bivectors consists of a set of two and three specific 
1-forms, respectively. Such 1-forms will determine 
codistributions which are integrable in the sense of Frobenius.
We will prove that the resulting bivectors are in fact Poisson.

\subsection{Some Poisson structures of rank two in $\R^4$\label{Pois_struc_R4}}

Consider the Euclidean space $\R^4$, with coordinates $(x_1,x_2,x_3,x_4)$. 
The equations of motion of the reduced non-holonomic systems encountered
in the examples are observed to have integrals of motion which are
related to the solutions of a system of differential equations of the type
\begin{eqnarray}
\frac{dx_3}{dx_1}=h_3(x_1,x_3,x_4)\,,\quad
\frac{dx_4}{dx_1}=h_4(x_1,x_3,x_4)\,,
\label{syst_first_ints}
\end{eqnarray}
where $h_3$, $h_4$, are two given (smooth) 
functions of their arguments, which do not include $x_2$. 
We consider the system (\ref{syst_first_ints}) as the Pfaffian system 
\lq $\th_1=0, \th_2=0$\rq, where the 1-forms $\th_1,\th_2$ in $\R^4$, 
are given by 
\begin{eqnarray}
\fl\quad\quad \th_1=-h_3(x_1,x_3,x_4)dx_1+dx_3\,,\quad
\th_2=-h_4(x_1,x_3,x_4)dx_1+dx_4\,.
\label{one_forms_first_ints}
\end{eqnarray}
These two 1-forms determine a codistribution integrable in the 
sense of Frobenius \cite{LibMarle87}, since there exist 
a set of four 1-forms $\D_i^j$ such that 
$d\th_i=\D_i^j\wedge\th_j$ for $i,j=1,2$.
For example, we can take 
\begin{equation}
\fl\quad\quad
\D_1^1=\pd{h_3}{x_3}dx_1\,,\quad
\D_1^2=\pd{h_3}{x_4}dx_1\,,\quad
\D_2^1=\pd{h_4}{x_3}dx_1\,,\quad
\D_2^2=\pd{h_4}{x_4}dx_1\,,
\label{one_forms_Frob}
\end{equation}
in order to satisfy the integrability condition. 
Thus, there will exist (locally) functions $c_1, c_2$
such that $\th_i=dc_i$, $i=1,2$. The subvarieties 
solution of the Pfaffian system \lq $\th_1=0, \th_2=0$\rq\ 
are defined by the equations $c_i=b_i$, 
where $b_i$ are constants, $i=1,2$.

More specifically, in the actual examples, the system 
(\ref{syst_first_ints}) takes the form of a non-autonomous 
first order system of linear differential equations
\begin{eqnarray}
\frac{dx_3}{dx_1}=a_{11}(x_1)x_3+a_{12}(x_1)x_4\,,\quad
\frac{dx_4}{dx_1}=a_{21}(x_1)x_3+a_{22}(x_1)x_4\,,	\nonumber
\end{eqnarray}
or, written in matrix form,
\begin{eqnarray}
\frac{d}{dx_1}\matriz{c}{x_3\\ x_4}=A(x_1)\matriz{c}{x_3\\ x_4}\,,
\label{matr_equation}
\end{eqnarray}
where 
\begin{eqnarray}
A(x_1)=
\matriz{cc}
{a_{11}(x_1) & a_{12}(x_1) \\ 
a_{21}(x_1) & a_{22}(x_1)}\,.	\nonumber
\end{eqnarray}
The previous functions $c_i$ can be identified with 
the initial conditions of the solution of (\ref{matr_equation}).
In fact, such a solution can be expressed as 
${\bf x}=g(x_1){\bf c}$, where ${\bf x}=(x_3, x_4)^T$, ${\bf c}=(c_1, c_2)^T$, 
and $g(x_1)$ is a $GL(2,\R)$-valued curve 
($SL(2,\R)$-valued curve if $\tr A(x_1)=0$ for all $x_1$), 
solution of the right-invariant matrix system (see, e.g., \cite{CarGraRam01,CarRam02})
\begin{equation}
\frac{dg}{dx_1}g^{-1}=A(x_1)\,.
\label{right_inv_matrix_syst_x4}
\end{equation} 
Then, ${\bf c}=g^{-1}(x_1){\bf x}$ gives the desired functions: with a 
slight abuse of notation, we have
$$
d{\bf c}=(dg^{-1}){\bf x}+g^{-1}d{\bf x}
=-g^{-1}dg g^{-1}{\bf x}+g^{-1}A{\bf x}\,dx_1
=-g^{-1}dg g^{-1}{\bf x}+g^{-1}dg g^{-1}{\bf x}=0\,,
$$
where we have used that $dg^{-1}=-g^{-1}dgg^{-1}$ and $A dx_1=dg g^{-1}$.  
However, note that the solution of (\ref{right_inv_matrix_syst_x4}) 
cannot be expressed in an explicit way in the general case, and therefore, 
the functions $c_1$, $c_2$ cannot be explicitly written in general.

Now, we impose that the 1-forms (\ref{one_forms_first_ints}) 
generate the kernel of the bivector in $\R^4$
\begin{equation}
\L=\sum_{1\leq i<j\leq 4}\L_{ij}\pd{}{x_i}\wedge\pd{}{x_j}\,.
\label{bivector_x4}
\end{equation}
The resulting bivectors will clearly have rank two. Moreover, they are 
Poisson, according to the following result
\begin{theorem}
Consider in $\R^4$ a bivector of type (\ref{bivector_x4}), such that 
$\L^\sharp(\th_1)=0$, $\L^\sharp(\th_2)=0$, where $\th_1$, $\th_2$ are given by
(\ref{one_forms_first_ints}). Then the bivector is of the form
$$
\L=-\L_{12}\,U\wedge V\,,
$$
where
\begin{equation}
U=\pd{}{x_2}\,,\quad V=\pd{}{x_1}+h_3\pd{}{x_3}+h_4\pd{}{x_4}\,,
\label{def_U_V_x4}
\end{equation} 
and $\L_{12}\in C^\infty(\R^4)$. Each of these bivectors is Poisson, 
and of rank two on points where $\L_{12}\neq 0$.
\label{propo_Pois_x4}
\end{theorem}

{\noindent{\it Proof}} 

The case of $\L_{12}=0$ is trivial. 
We will assume $\L_{12}\neq 0$ in the domain of interest. 
Take $\L$ and $\th_1$, $\th_2$, as stated. The conditions 
$\L^\sharp(\th_1)=0$, $\L^\sharp(\th_2)=0$ give rise to an algebraic 
system for the six independent functions $\L_{ij}$, which can be easily solved 
for five of them, in terms of the remaining one and the functions
entering into the 1-forms. We choose $\L_{12}$ to be the 
undetermined function. Then the solution reads
$$
\L_{13}=\L_{14}=\L_{34}=0\,,\quad \L_{23}=-\L_{12}h_3\,,\quad\L_{24}=-\L_{12}h_4\,,
$$
thus the resulting bivectors are as claimed. 
To see that each of them is Poisson, consider the bivector 
of the family with $\L_{12}=-1$, i.e., $U\wedge V$.
This bivector is of the form given in Example~1, and $[U,V]=0$,
thus $U\wedge V$ is Poisson. It is moreover of rank two, 
therefore by Proposition~\ref{prop_conf}, the claim follows. 
  
\medskip
{\noindent {\bf Remark}} 
Note that the vector fields $U,V$ of the previous Theorem
satisfy $\th_i(U)=\th_i(V)=0$, $i=1,2$, which in principle
might seem a stronger condition than that the bivector 
(\ref{bivector_x4}) annihilates the 1-forms $\th_1,\th_2$.

\medskip
Now, given a (Hamiltonian) function $H\in C^\infty(\R^4)$, 
the Hamiltonian vector field $X_H$ with respect to a Poisson 
structure of the family described on Theorem~\ref{propo_Pois_x4}
takes the form
\begin{eqnarray}
X_H=\L^\sharp(dH)=\L_{12}\left[(VH)U-(UH)V\right]\,,
\label{ham_vf_x4}
\end{eqnarray}
where $U$ and $V$ are given by (\ref{def_U_V_x4}). 
Obviously, $H$ is a first integral of $X_H$, since $X_HH=\L(dH,dH)=0$.
Other two first integrals are the functions $c_i$ such that $dc_i=\th_i$, 
since by construction $X_H(c_i)=\L(dH,dc_i)=\L(dH,\th_i)=0$, $i=1,2$.
These two first integrals are common to all Hamiltonian vector 
fields of type (\ref{ham_vf_x4}). 

On the other hand, given a specific vector field $X$ in $\R^4$, which is 
recognized to be of the form (\ref{ham_vf_x4}), it could be regarded as a 
Hamiltonian vector field with respect to \emph{one specific} 
Poisson structure of the family described in Theorem~\ref{propo_Pois_x4}.   

\subsection{Some Poisson structures of rank two in $\R^5$\label{Pois_struc_R5}}

We will treat in this Section analogous questions to that of the 
previous Section, but now in the Euclidean space $\R^5$, 
with coordinates $(x_1,x_2,x_3,x_4,x_5)$. 

The motivation is that typically, the reduced orbit spaces for 
the non-holonomic problems of interest, are semialgebraic varieties 
of $\R^5$, essentially determined by the zero level set of a 
function $\phi\in C^\infty(\R^5)$, quadratic in its arguments,
which are moreover subject to certain constraints. 
More specifically, in the examples it will have the form
$\phi(x)=0$, with $\phi(x)=x_2^2+x_3^2-(1-x_1^2)x_5$, $|x_1|\leq 1$, and $x_5\geq 0$,  
or with $\phi(x)=x_2^2+x_3^2-4\,x_1 x_5$, $x_1\geq 0$, and $x_5\geq 0$.   
However, for what follows $\phi$ can be in principle 
any differentiable function in $\R^5$. 

We will consider then the Pfaffian 
system \lq $\th_0=0, \th_1=0, \th_2=0$\rq, where $\th_0=d\phi$ 
and $\th_1,\th_2$ are 1-forms in $\R^5$ whose coordinate expression 
is again (\ref{one_forms_first_ints}). These three 1-forms also
determine determine a codistribution integrable in the 
sense of Frobenius in $\R^5$, because we have again 
$d\th_i=\D_i^j\wedge\th_j$ with (\ref{one_forms_Frob}), $i,j=1,2,$
and $d\th_0=d^2\phi=0$.

We impose now that $\ker\,\L^\sharp={\rm span}\{\th_0,\th_1,\th_2\}$,
where $\L$ is the bivector in (some open set of) $\R^5$
\begin{equation}
\L=\sum_{1\leq i<j\leq 5}\L_{ij}\pd{}{x_i}\wedge\pd{}{x_j}\,.
\label{bivector_x5}
\end{equation}
The resulting bivectors are again generically of rank two and Poisson, as follows 
\begin{theorem}
Consider in $\R^5$ a bivector of type (\ref{bivector_x5}), such that 
$\L^\sharp(\th_0)=0$, $\L^\sharp(\th_1)=0$ and 
$\L^\sharp(\th_2)=0$, where $\th_0=d\phi$, and $\th_1$, $\th_2$ 
are given by (\ref{one_forms_first_ints}). 
Then the bivector is of the form
\begin{equation}
\L=f[(Z\phi)U\wedge V+Y\wedge Z]\,,
\label{Pois_biv_x5}
\end{equation}
where
\begin{eqnarray}
&& U=\pd{}{x_2}\,,\quad V=\pd{}{x_1}+h_3\pd{}{x_3}+h_4\pd{}{x_4}\,,\quad Z=\pd{}{x_5}\,,
\label{def_U_V_Z_x5}	\\
&& Y=(U\phi)V-(V\phi)U\,,
\label{def_Y_x5}	
\end{eqnarray}
and $f\in C^\infty(\R^5)$. 
Each of these bivectors is Poisson, and of rank two on points where $f\neq 0$.
\label{propo_Pois_x5}
\end{theorem}

{\noindent{\it Proof}} 

Once more, the case of $f=0$ is trivial,
thus we will assume again that $f\neq 0$ in the domain of interest.  
Take $\L$, $\th_0$, $\th_1$ and $\th_2$ as stated. 
The idea of the proof is similar to that of Theorem~\ref{propo_Pois_x4}.
First of all, since the kernel of $\L^\sharp$ has generically 
dimension three, then the rank of $\L^\sharp$ is two. 
The conditions $\L^\sharp(\th_0)=0$, $\L^\sharp(\th_1)=0$ and $\L^\sharp(\th_2)=0$
give rise again to an algebraic system for the functions $\L_{ij}$, out of
which all can be solved for except one of them, namely $\L_{12}$, which we will write
as $-(\partial{\phi}/\partial{x_5})f$.
The solution then reads
\begin{eqnarray}
&& \L_{13}=\L_{14}=\L_{34}=0\,,\quad \L_{23}=f h_3\pd{\phi}{x_5}\,,
\quad\L_{24}=f h_4\pd{\phi}{x_5}\,,				\nonumber\\
&& \L_{15}=f \pd{\phi}{x_2}\,,\quad
\L_{35}=f h_3\pd{\phi}{x_2}\,,					\nonumber\\
&& \L_{45}=f h_4\pd{\phi}{x_2}\,,
\quad 
\L_{25}=-f
\left(
\pd{\phi}{x_1}+h_3\pd{\phi}{x_3}+h_4\pd{\phi}{x_4}
\right)\,,							\nonumber
\end{eqnarray}
thus the resulting bivectors take the stated form.
To see that each of them is Poisson, consider the bivector of 
the family with $f=1$, i.e., 
$\L_0=\overline{U}\wedge V+Y\wedge Z$, where $\overline{U}=(Z\phi)U$. 
We have to show that the Schouten--Nijenhuis bracket of $\L_0$ 
with itself vanish, i.e., $[\L_0,\L_0]=0$. 
By linearity and using the first property of (\ref{props_Scho}) we have
\begin{eqnarray}
[\L_0,\L_0]&=&[\overline U\wedge V,\overline U\wedge V]+2\,[\overline U\wedge V,Y\wedge Z]
+[Y\wedge Z,Y\wedge Z] 
\nonumber
\end{eqnarray}
By Example~1 we know that $[\overline U\wedge V,\overline U\wedge V]
=2\,\overline U\wedge V\wedge[\overline U,V]$ 
and analogously, $[Y\wedge Z,Y\wedge Z]=2\,Y\wedge Z\wedge[Y,Z]$. 
Now, using again the second and third properties of (\ref{props_Scho}) we can write
\begin{eqnarray}
[\overline U\wedge V,Y\wedge Z]&=&V\wedge Z\wedge[\overline U,Y]
-\overline U\wedge Z\wedge[V,Y]	\nonumber\\
& &+Y\wedge V\wedge[\overline U,Z]-Y\wedge \overline U\wedge[V,Z]			
\nonumber
\end{eqnarray}
We have to calculate now some Lie brackets. We have $[U,V]=[V,Z]=[U,Z]=0$ but
\begin{eqnarray}
&& \fl\quad\quad [{\overline U},V]
=-[V(Z\phi)]U\,,\quad [Y,Z]=-[Z(U\phi)]V+[Z(V\phi)]U		\nonumber\\
&& \fl\quad\quad [{\overline U},Y]=(Z\phi)[U(U\phi)]V
-\{(Z\phi)[U(V\phi)]+(U\phi)[V(Z\phi)]-(V\phi)[U(Z\phi)]\}U	\nonumber\\
&& \fl\quad\quad [V,Y]=[V(U\phi)]V-[V(V\phi)]U\,,
\quad[{\overline U},Z]=-[Z(Z\phi)]U				\nonumber
\end{eqnarray}
Then, summing up, we have
\begin{eqnarray}
\fl\quad\quad  [\L_0,\L_0]=2\, U\wedge V\wedge Z
\{(Z\phi)([V,U]\phi)+(U\phi)([Z,V]\phi)+(V\phi)([U,Z]\phi)\}=0\,.\nonumber
\end{eqnarray}
Since the rank of any of the $\L$, 
and in particular $\L_0$, is two, 
applying Proposition~\ref{prop_conf} ends the proof.

\medskip
{\noindent {\bf Remark}} 
Note that the vector fields $U,V,Y$ and $Z$ of Theorem~\ref{propo_Pois_x5}
satisfy $\th_i(U)=\th_i(V)=\th_i(Y)=\th_i(Z)=0$, $i=1,2$, 
$\th_0(Y)=Y(\phi)=0$ and $(U\wedge V)\phi-Y=0$. These requirements
might seem \emph{a priori} to be stronger conditions to that 
imposed in the Theorem. 

\medskip 
If we are given now a (Hamiltonian) function $H\in C^\infty(\R^5)$, 
the Hamiltonian vector field $X_H$ with respect to a Poisson 
structure of the family described in Theorem~\ref{propo_Pois_x5} reads, 
using (\ref{def_Y_x5}),
\begin{eqnarray}
&&X_H=\L^\sharp(dH)=
f
\big\{
[(ZH)(V\phi)-(Z\phi)(VH)]\,U	\label{ham_vf_x5}\\
&&\quad+[(Z\phi)(UH)-(ZH)(U\phi)]\,V
+[(U\phi)(VH)-(V\phi)(U H)]\,Z
\big\}\,,
\nonumber
\end{eqnarray}
where $U$, $V$ and $Z$ are given by (\ref{def_U_V_Z_x5}). 
By construction $H$ is a first integral of $X_H$. Other first integrals
are the functions $c_i$ such that $dc_i=\th_i$, as in the previous Section.
These two first integrals are common to all Hamiltonian vector 
fields of type (\ref{ham_vf_x5}).

However, given a specific vector field $X_H$ of type 
(\ref{ham_vf_x5}), it fixes the specific function $f$ and therefore the specific
Poisson bivector of the family (\ref{Pois_biv_x5}) with respect to which $X_H$ 
is Hamiltonian. 

\section{Examples\label{sect_of_examples}}

In this Section we will show how the preceding results can be 
directly applied in the cases of reduced systems corresponding to
specific examples of non-holonomic systems, i.e., the rolling disk, 
the Routh's sphere, the ball rolling on a surface of revolution 
and its special case of a ball rolling inside a cylinder.

\subsection{The rolling disk\label{sect_rolling_disk}}

For this example we will follow the treatment and use some of 
the results of \cite{CusHerKem96}, see details therein. 
This problem has been treated as well, e.g., in 
\cite{BatGraMac96,BorMam02c,CusKemSniBat95,NeiFuf72,Rou60}.
Consider a homogeneous disk, which rolls without slipping on 
a horizontal plane under the influence of a vertical gravitational 
field of strenght $g$.
The resulting non-holonomic system has two evident symmetry groups.
One is the symmetry group $E(2)$ consisting of translations in 
the horizontal plane and rotations about the vertical axis, and 
the second is the $S^1$ symmetry consisting of rotations about 
the principal axis perpendicular to the plane of the disk. 

After these two symmetries have been reduced out, in particular by 
using invariant theory for the reduction of the $S^1$ symmetry,
it is obtained a system giving the evolution on the reduced 
orbit space, which is a semialgebraic variety of $\R^5$.
In particular, the system can be restricted to a smooth open 
subset as it has been done in \cite{CusHerKem96}. 

Thus, consider a reference 
homogeneous disk of radius $r$ and mass $m$, lying flat in a fixed
reference frame with center of mass at the origin. 
The position of 
the moving disk is given by transforming the position 
of the reference disk by means of a translation $a$
(e.g., of the center of mass) and a rotation $A$.
The tensor of inertia $I$ with respect to the principal axes of the disk
is diagonal, $I={\rm diag}(I_1,I_1,I_3)$. Let us call $e_3$ the vertical 
unitary vector in the fixed frame of reference. We define the unitary 
vector $u$ with respect that frame as the pre-image of $-e_3$ under the 
rotation $A$, $u=-A^{-1}e_3$. The vector $s$ in the fixed disk, rotated 
by $A$ gives the vector in the moving disk pointing {}from the center of mass 
to the point of contact of the moving disk with the horizontal plane. 
If we denote $\hat u=u-\langle u,e_3\rangle e_3$, the relation between
$s$ and $u$ is $s=r\,{\hat u}/{|\hat u|}$. 
We denote by $(\om_1,\om_2,\om_3)$ the components of the angular 
velocity vector $\om$ of the disk,

Following \cite{CusHerKem96}, after the mentioned symmetry 
group $E(2)$ is reduced out, the equations of motion read
\begin{eqnarray}
&&\frac{d(I\om)}{dt}=I\om\times\om-m r^2\frac{d\om}{dt}
+m\left\langle\frac{d\om}{dt},s\right\rangle s
+m\langle s,\om\rangle\frac{ds}{dt}			\nonumber\\
&&\quad\quad\quad\quad
+m\langle\om,s\rangle(\om\times s)-mg\,(u\times s) \label{eqs_mot_roll_disk}\\
&& \frac{du}{dt}=u\times\om \nonumber
\end{eqnarray}
which have a first integral given by the total energy of the disk
\begin{eqnarray}
&&
H=\frac12\langle I\om,\om\rangle+\frac12\langle\om\times s,\om\times s\rangle
+mg\,\langle s,u\rangle\,.					\label{ener_mot_roll_disk}
\end{eqnarray}
The second of Eqs. (\ref{eqs_mot_roll_disk}) expresses the non-holonomic
constraint of rolling without slipping, i.e., instantaneous velocity 
of the point of contact equal to zero.

We recall briefly now how the further reduction of the $S^1$ symmetry 
is performed.
Let us denote by $(u_1,u_2,u_3)$ the components of $u$.
The $S^1$ symmetry action consists of rotating both vectors $u$ and $\om$
simultaneously as mentioned, and it is not a free action since the 
isotropy subgroup of pairs $((0,0,\pm 1),(0,0,\om_3))$ is $S^1$.
Thus, we will use invariant theory in order to perform the reduction.
A set of invariants for this action is easily constructed \cite{CusHerKem96}:
\begin{eqnarray}
&& \s_1=u_3\,,\quad\s_2=u_2\om_1-u_1\om_2\,,\quad\s_3=u_1\om_1+u_2\om_2\,, \nonumber\\
&& \s_4=\om_3\,,\quad\s_5=\om_1^2+\om_2^2\,,\quad\s_6=u_1^2+u_2^2\,, \label{inv_sigma}
\end{eqnarray}
with the relations 
\begin{equation}
\s_2^2+\s_3^2=\s_5\s_6\,,\quad \s_5\geq 0\,,\quad \s_6\geq 0\,.
\label{rels_inv_sigma}
\end{equation}
Since $u$ is a unitary vector, we have that $\s_6+\s_1^2=1$ and $|\s_1|\leq 1$, 
thus the completely reduced orbit space $M$ is the semialgebraic variety of $\R^5$ 
\begin{equation}
M=\{(\s_1,\dots,\s_5)\in\R^5\,\,|\,\,\phi(\s)=0,\ |\s_1|\leq 1,\ \s_5\geq 0\}\,,
\label{semialg_var_1}
\end{equation}
where $\phi\in C^\infty(\R^5)$ is the polynomial function 
$\phi(\s)=\s_2^2+\s_3^2-(1-\s_1^2)\s_5$. However, $M$ is \emph{not} 
a smooth submanifold of $\R^5$. The singular points of $M$ are 
\begin{equation}
\Pi_{\pm}=\{(\pm 1,0,0,\s_4,\s_5)\in\R^5\,\,|\,\,\s_4\in\R,\ \s_5\geq 0\}\,.
\label{sing_semialg_var_1}
\end{equation}
The non-smoothness of $M$ is due to the fact that the $S^1$ action is not free, 
see \cite{Cus98}. 

The somehow redundant variables $(\s_1,\s_2,\s_3,\s_4,\s_5)$ therefore parametrize 
the reduced orbit space $M$. The induced system {}from (\ref{eqs_mot_roll_disk})
will be written in terms of the orbit variables: simply calculating
their time-derivatives, using the equations of motion (\ref{eqs_mot_roll_disk})
and that $I_1=\frac14mr^2$ and $I_3=\frac12mr^2$, we arrive to the
following system
\begin{eqnarray}
&&\dot\s_1=\s_2						\nonumber\\
&&\dot\s_2=\frac 6 5\,\s_3\s_4
-\s_1\s_5+\frac45\frac{\s_1 \s_3^2}{1-\s_1^2}+\l\s_1\sqrt{1-\s_1^2}	\nonumber\\
&&\dot\s_3=-2 \s_2\s_4					\label{sist_sig_5_roll_disk}\\
&&\dot\s_4=-\frac 23 \frac{1}{1-\s_1^2}\,\s_2\s_3	\nonumber\\
&&\dot\s_5=2 \s_2\left(\frac{\l \s_1}{\sqrt{1-\s_1^2}}
+\frac45\frac{\s_1\s_3^2}{(1-\s_1^2)^2}
-\frac45\frac{\s_3\s_4}{1-\s_1^2}
\right)\,,								\nonumber
\end{eqnarray} 
where $\l=\frac{4}{5}\frac g r$ and the dot means derivative with respect to time.  
The reduced energy, obtained {}from (\ref{ener_mot_roll_disk}), reads 
\begin{equation}
E=\frac{\s_5}2+\frac34\s_4^2-\frac25\frac{\s_3^2}{1-\s_1^2}+\l\sqrt{1-\s_1^2}\,.
\label{ener_sig_5_roll_disk}
\end{equation}
Although in principle the expressions 
(\ref{sist_sig_5_roll_disk}) and (\ref{ener_sig_5_roll_disk}) are only defined on $M$,
their right hand sides make sense for 
${\cal D}
=\R^5\backslash(\{(\pm1,\s_2,\s_3,\s_4,\s_5)\ |\ \s_2\s_3\neq 0\}
\cup\{(\s_1,\s_2,\s_3,\s_4,\s_5)\ | \ |\s_1|>1\})$,
so we will consider this extended domain for the vector field $X$
whose integral curves are given by (\ref{sist_sig_5_roll_disk}) and
the reduced energy function $E$.

However, if we restrict ourselves to the original domain $M$, 
and morever to points with $|\s_1|<1$, we can define a smooth
open dense subset $\overline{M}\subset M$ given by
\begin{equation}
\overline{M}
=\left\{
(\s_1,\s_2,\s_3,\s_4,\s_5)\in\R^5\,\,\big|\,\,\s_5=\frac{\s_2^2+\s_3^2}{1-\s_1^2},\ |\s_1|<1
\right\}\,,
\label{smooth_subman}
\end{equation}
diffeomorphic to $\R^4$ \cite{CusHerKem96}. The induced vector field
$\overline{X}$ and energy $\overline{E}$ on $\overline{M}$ can be easily 
found {}from (\ref{sist_sig_5_roll_disk}) and (\ref{ener_sig_5_roll_disk}) by
solving for $\s_5$. The integral curves of $\overline{X}$ are the solutions of
the system
\begin{eqnarray}
&&\dot\s_1=\s_2						\nonumber\\
&&\dot\s_2=\frac 6 5\,\s_3\s_4
-\frac{\s_1}{1-\s_1^2}\,\s_2^2
-\frac 1 5\frac{\s_1}{1-\s_1^2}\,\s_3^2+\l\s_1\sqrt{1-\s_1^2}	\label{sist_rd_sig4}\\
&&\dot\s_3=-2 \s_2\s_4					\nonumber\\
&&\dot\s_4=-\frac 23 \frac{1}{1-\s_1^2}\,\s_2\s_3\,,	\nonumber
\end{eqnarray} 
meanwhile
\begin{equation}
\overline{E}=\frac 1 2 \frac{\s_2^2}{1-\s_1^2}
+\frac 1{10}\frac{\s_3^2}{1-\s_1^2}+\frac 3 5\,\s_4^2+\l\sqrt{1-\s_1^2}\,.
\label{ener_rd_sig4}
\end{equation}
These expressions are Eqs. (18) and (19) of \cite{CusHerKem96}, respectively.

The reduced vector field $X$ satisfies $X(E)=0$ as well as $X(\phi)=0$
in ${\cal D}$, meanwhile $\overline{X}(\overline{E})=0$ in $\overline{M}$.
In addition, $X$ has a family of equilibrium points belonging to the singular
set $\Pi_\pm$, called \emph{singular equilibria}, given by 
$\{(\pm 1,0,0,\s_4,0)\ |\ \s_4\in\R\}$, and a family 
of \emph{regular equilibria} given by the set of constants 
$$
\left\{(\s_{10},0,\s_{30},\s_{40},\s_{50})\in {\cal D}\ \bigg|\ \frac 6 5\,\s_{30}\s_{40}
-\s_{10}\s_{50}
+\frac45\frac{\s_{10} \s_{30}^2}{1-\s_{10}^2}+\l\,\s_{10}\sqrt{1-\s_{10}^2}=0
\right\}\,.
$$
These regular equilibria, in the original system, correspond to periodic motions
of the disk in which the point of contact describes a circle and the center of
mass stands at constant height. These motions are contained in the set of 
\emph{steady motions} of the rolling disk, according to Routh's terminology 
\cite{NeiFuf72,Rou60}. They have received an extensive treatment 
in \cite{CusHerKem96}, although by using the system (\ref{sist_rd_sig4}).

Now, both of the systems (\ref{sist_sig_5_roll_disk}) and (\ref{sist_rd_sig4}) 
admit two first integrals related to the solutions (in the sense explained 
in Section~\ref{Pois_struc_R4}) of the non-autonomous linear system 
\begin{eqnarray}
\frac{d\s_3}{d\s_1}=-2\s_4\,,\quad\frac{d\s_4}{d\s_1}=-\frac23\frac{\s_3}{1-\s_1^2}\,,
\end{eqnarray}
which can be written in matrix form as
\begin{equation}
\frac{d}{d\s_1}\matriz{c}{\s_3\\ \s_4}
=\matriz{cc}{0 & -2 \\ -\frac 2 3 \frac 1{1-\s_1^2} & 0}\matriz{c}{\s_3\\ \s_4}\,.
\label{Chap_eqs}
\end{equation}
This equation is the same as Eq. (69) of \cite{CusHerKem96}, 
where its solutions have been studied in great detail, 
including their asymptotic behaviour. 

However, the important point for us is that the systems 
(\ref{sist_sig_5_roll_disk}) and (\ref{sist_rd_sig4}) 
are good candidates to be formulated as Hamiltonian systems with 
respect to Poisson structures of the type described 
in Theorems~\ref{propo_Pois_x5} and~\ref{propo_Pois_x4}, respectively. 
Let $\th_0=d\phi$ and $\th_1$, $\th_2$ be the 1-forms, defined in $\overline{M}$
(resp. ${\cal D}$) by 
\begin{eqnarray}
&& \th_1=2 \s_4\,d\s_1+d\s_3\,,
\quad \th_2=\frac23\frac{\s_3}{1-\s_1^2}\, d\s_1+d\s_4\,. 	\nonumber
\end{eqnarray}
Applying the results of Sections~\ref{Pois_struc_R4} and~\ref{Pois_struc_R5}
to these 1-forms, we have 

\begin{proposition}
The bivectors of the form $\overline{\L}=-\L_{12}\,U\wedge V$, 
defined in $\overline{M}$, where 
\begin{eqnarray}
U=\pd{}{\s_2}\,,\quad V=\pd{}{\s_1}-2 \s_4\pd{}{\s_3}
-\frac23\frac{\s_3}{1-\s_1^2}\pd{}{\s_4}\,,
\label{def_U_V_Z_x5_roll_disk}	
\end{eqnarray}
and $\L_{12}\in C^\infty(\overline{M})$ is a non-vanishing function, 
are Poisson tensors of rank two in $\overline{M}$. 

The vector field $\overline{X}$ in $\overline{M}$, 
whose integral curves are the solutions of (\ref{sist_rd_sig4}),
is a Hamiltonian vector field with respect to the Poisson bivector
$\overline{\L}$ with the specific function $\L_{12}=1-\s_1^2$ and 
Hamiltonian function $\overline{E}$ given by (\ref{ener_rd_sig4}), 
i.e., $\overline{X}=\overline{\L}^\sharp(d\overline{E})$ in $\overline{M}$.
\label{prop_Lam_s_roll_disk_s4}
\end{proposition}

\begin{proposition}
The bivectors $\L=f[(Z\phi)U\wedge V+Y\wedge Z]$, 
defined in ${\cal D}\subset\R^5$, where $U$ and $V$ are given 
by (\ref{def_U_V_Z_x5_roll_disk}), $Z=\partial{}/\partial{\s_5}$, 
$Y=(U\phi)V-(V\phi)U$, and $f\in C^\infty({\cal D})$ 
is a non-vanishing function, are Poisson tensors of rank two in ${\cal D}$,
except in the set of singular equilibria, where they vanish.

The vector field $X$ in ${\cal D}$, whose integral 
curves are the solutions of (\ref{sist_sig_5_roll_disk}),  
is a Hamiltonian vector field with respect to the Poisson bivector
$\L$ with the specific function $f=1$ and 
Hamiltonian function $E$ given by (\ref{ener_sig_5_roll_disk}), 
i.e., $X=\L^\sharp(dE)$ in ${\cal D}$.
\end{proposition}

Both Propositions can be proved by direct computations. 

\medskip
The Poisson Hamiltonian structure of the systems
(\ref{sist_sig_5_roll_disk}) and (\ref{sist_rd_sig4}) 
could be used to have an interpretation of their geometry. 
For example, the invariant submanifolds mentioned in the analysis 
of the reduced vector field (\ref{sist_rd_sig4}) in \cite{CusHerKem96}, 
could be understood as the symplectic leaves of the rank-two Poisson structure(s) 
$\overline{\L}$ of Proposition~\ref{prop_Lam_s_roll_disk_s4}.

\subsection{Routh's sphere\label{sect_Routh_sphere}}

For this example we will follow the treatment and use some of 
the results of \cite{Cus98}, see details therein.
This problem has been treated as well, e.g., in 
\cite{BatGraMac96,BorMam02c,EbeSch95,NeiFuf72,Rou60}.
Consider a sphere of mass $m$ and radius $r$ with its center of mass 
at a distance $\a$ ($0<\a<r$) {}from its geometric center. 
The line joining both centers is a principal axis of inertia, 
with associated moment of inertia $I_3$. 
Any axis orthogonal to the previous, passing though the geometric center, has
an associated moment of inertia $I_1$. This sphere is supposed to roll on a 
horizontal plane under the influence of a vertical gravitational field of strenght $g$. 
The resulting non-holonomic system has as well two symmetry groups. 
One is again the group $E(2)$ consisting of translations in 
the horizontal plane and rotations about the vertical axis. 
The other is the $S^1$ symmetry consisting of rotations about the 
principal axis of inertia which joins the center of mass and 
the geometric center of the ball. 
 
Again, after these symmetries have been reduced out by a 
similar procedure to that of the rolling disk, 
it is obtained a system giving the evolution on the reduced 
orbit space, which coincides with that of the rolling disk.

Therefore, let us consider a reference ball as the one described, 
with the geometric center at the origin, and the center of mass at the point
$-\a e_3$, where $e_3$ denotes the vertical unitary vector in this fixed frame.
The position and attitude of the moving ball is given
by transforming the position of the reference ball by means of a 
translation $a$ (e.g., of the center of mass) and a rotation $A$.
We denote by $s$ the vector in the fixed sphere such that rotated by $A$
gives the vector in the moving sphere pointing {}from the center of mass 
to the point of contact. The unitary vector $u$ in the fixed frame is the 
pre-image of $-e_3$ under the rotation $A$, $u=-A^{-1}e_3$. The relation
between $u$ and $s$ is $a_3=\langle s,u\rangle$.
The components of the angular velocity $\om$ of the ball will be 
denoted by $(\om_1,\om_2,\om_3)$. 

Following \cite{Cus98}, after the reduction of the mentioned 
$E(2)$ symmetry, the equations of motion read
\begin{eqnarray}
&&\frac{d}{dt}(I\om+m s\times(\om\times s))=I\om\times\om
+m\frac{ds}{dt}\times(\om\times s)			\nonumber\\
&& \quad\quad\quad\quad\quad\quad\quad\quad\quad\quad\quad\quad
+m\langle \om,s \rangle(\om\times s)+mg\,(u\times s)
\label{eqs_mot_Routh_esf}\\
&& \frac{du}{dt}=u\times\om \nonumber
\end{eqnarray}
which have a first integral given by the total energy of the ball
\begin{eqnarray}
&&
H=\frac12\langle I\om,\om\rangle+\frac12\langle\om\times s,\om\times s\rangle
+mg\,\langle s,u\rangle\,.					\label{ener_mot_Routh_esf}
\end{eqnarray}
The second of Eqs.~(\ref{eqs_mot_Routh_esf}) expresses again the non-holonomic
constraint of rolling without slipping.

Now, the reduction of the $S^1$ symmetry is performed in an analogous way as
in the case of the rolling disk, see Section~\ref{sect_rolling_disk}, where
$(u_1,u_2,u_3)$ denote as well the components of $u$. The $S^1$ action consists 
of rotating both vectors $u$, $\om$ simultaneously, with respect to 
the principal axis joining the geometric and mass centers. 
This action is not free, since $S^1$ leaves invariant 
pairs of points of the form $((0,0,\pm 1),(0,0,\om_3))$.
The corresponding set of invariants is again (\ref{inv_sigma}) with the 
relations (\ref{rels_inv_sigma}). Thus, the reduced orbit space $M$ 
is the semialgebraic variety of $\R^5$ described in the previous example 
of the rolling disk, with the same notations.

However, the reduced system reads now, using (\ref{eqs_mot_Routh_esf}), 
\begin{eqnarray}
\quad\quad\ \ \dot\s_1&=&\s_2						\nonumber\\
T(\s_1) \dot\s_2&=&
(I_3+m r^2+m r\a\s_1)\s_3\s_4-m g \a(1-\s_1^2)		\nonumber\\
& &\ \ 
-\s_5(m r \a+(I_1+m \a^2+m r^2)\s_1+m r \a\s_1^2)	\nonumber\\
\quad\quad\ \ \dot\s_3&=&-I_3\frac{\s_2\s_4}{P(\s_1)}(I_3+m r^2+m r \a\s_1)	
							\label{sist_sig_5_routh_esf}\\
\quad\quad\ \ \dot\s_4&=&-m r\frac{\s_2\s_4}{P(\s_1)}(I_3\a+r(I_3-I_1)\s_1)\nonumber\\
T(\s_1)\dot\s_5&=&
-2 m r \a\s_2\s_5-2 m g\a\s_2				\nonumber\\
&&\quad -2 m r^2(I_3-I_1)
\frac{I_3+m r^2+m r\a\s_1}{P(\s_1)}\s_2\s_3\s_4\,,	\nonumber
\end{eqnarray} 
where 
$P(\s_1)=I_1 I_3+m r^2 I_1(1-\s_1^2)+m I_3(\a+r \s_1)^2$ and 
$T(\s_1)=I_1+m r^2+m\a^2+2 m r \a\s_1$. The reduced energy is
\begin{eqnarray}
\fl \quad E=\frac12(T(\s_1)\s_5+(I_3+mr^2)\s_4^2-mr^2(\s_3+\s_1\s_4)^2)
+m\a(g\s_1-r\s_3\s_4)\,.
\label{ener_sig_5_routh_esf}
\end{eqnarray}
These are eqs. (23), (24) and (25) in \cite{Cus98}.
In this case the expressions (\ref{sist_sig_5_routh_esf}) 
and (\ref{ener_sig_5_routh_esf}) make sense for all ${\cal D}=\R^5$. 
We will consider this extended domain for the vector field $X$ whose 
integral curves are given by (\ref{sist_sig_5_routh_esf}) and for 
the reduced energy function $E$.

Restricting ourselves to points in $M$ with $|\s_1|<1$,
we find again the smooth submanifold $\overline{M}\subset M$ 
given by (\ref{smooth_subman}). The integral curves of the 
projected vector field $\overline{X}$ are the solutions 
of the system (\cite{Cus98}, eq. (38))
\begin{eqnarray}
\quad\quad\ \ \dot\s_1&=&\s_2						\nonumber\\
T(\s_1) \dot\s_2&=&
(I_3+m r^2+m r\a\s_1)\s_3\s_4-m g \a(1-\s_1^2)		\nonumber\\
& &\ \ 
-\frac{\s_2^2+\s_3^2}{1-\s_1^2}(m r \a+(I_1+m \a^2+m r^2)\s_1+m r \a\s_1^2)	
							\label{sist_resf_sig4}\\
\quad\quad\ \ \dot\s_3&=&-I_3\frac{\s_2\s_4}{P(\s_1)}(I_3+m r^2+m r \a\s_1)	
							\nonumber\\
\quad\quad\ \ \dot\s_4&=&-m r\frac{\s_2\s_4}{P(\s_1)}(I_3\a+r(I_3-I_1)\s_1)
							\nonumber
\end{eqnarray} 
and the restricted reduced energy $\overline{E}$ is 
\begin{eqnarray}
\fl
\overline{E}=\frac12\left(T(\s_1)\frac{\s_2^2+\s_3^2}{1-\s_1^2}
+(I_3+mr^2)\s_4^2-mr^2(\s_3+\s_1\s_4)^2\right)
+m\a(g\s_1-r\s_3\s_4)\,.	
\label{ener_resf_sig4}
\end{eqnarray}

The reduced vector field $X$ satisfies $X(E)=0$ and $X(\phi)=0$ 
in ${\cal D}$, and $\overline{X}(\overline{E})=0$ in $\overline{M}$.
Moreover, $X$ has a family of singular equilibrium points, belonging 
to the singular set $\Pi_\pm$, given by $\{(\pm 1,0,0,\s_4,0)\ |\ \s_4\in\R\}$, 
which physically correspond to the spinning of the ball about 
its symmetry axis when it is vertical 
(then the reduced energy becomes $\frac12I_3\s_4^2\pm mg\a$).
It has as well a family of regular equilibria given by the set of constants 
$$
\{(\s_{10},0,\s_{30},\s_{40},\s_{50})\in \R^5\ |\ b(\s_{10},\s_{30},\s_{40},\s_{50})=0
\}\,,
$$
where $b(\s_{1},\s_{3},\s_{4},\s_{5})=(I_3+m r^2+m r\a\s_{1})\s_{3}\s_{4}
-m g \a(1-\s_{1}^2)-\s_{5}(m r \a+(I_1+m \a^2+m r^2)\s_{1}+m r \a\s_{1}^2)$.
These regular equilibria, in the original system, correspond to periodic motions
of the ball in which the point of contact describes a circle and the center of 
mass stands at constant height.

In this case, both of the systems (\ref{sist_sig_5_routh_esf}) 
and (\ref{sist_resf_sig4}) admit two first integrals related to 
the solutions (in the sense of Section~\ref{Pois_struc_R4}) of 
the non-autonomous linear system 
\begin{eqnarray}
\fl\quad\quad
\frac{d\s_3}{d\s_1}=-\frac{I_3(I_3+m r(r+\a\s_1))\s_4}{P(\s_1)}\,,	
\quad
\frac{d\s_4}{d\s_1}=\frac{m r(I_1 r\s_1-I_3(\a+r \s_1))\s_4}{P(\s_1)}\,.\label{sys_chap_rs}
\end{eqnarray}
Thus, the mentioned systems are other good candidates on which to apply 
the Poisson approach of Section~\ref{Pois_struc_R4_R5}.
Let $\th_0=d\phi$ and $\th_1$, $\th_2$ be the 1-forms, defined in $\overline{M}$ 
(resp. ${\cal D}$) by 
\begin{eqnarray}
&& \th_1=\frac{I_3(I_3+m r(r+\a\s_1))\s_4}{P(\s_1)}d\s_1+d\s_3\,,	\nonumber\\
&& \th_2=-\frac{m r(I_1 r\s_1-I_3(\a+r \s_1))\s_4}{P(\s_1)}d\s_1+d\s_4\,.
									\nonumber
\end{eqnarray}
We have the following results:
\begin{proposition}
The bivectors of the form $\overline{\L}=-\L_{12}\,U\wedge V$, 
defined in $\overline{M}$, where $U=\partial/\partial{\s_2}$,
\begin{eqnarray}
\fl\quad V=\pd{}{\s_1}
-\frac{I_3(I_3+m r(r+\a\s_1))\s_4}{P(\s_1)}\pd{}{\s_3}
+\frac{m r(I_1 r\s_1-I_3(\a+r \s_1))\s_4}{P(\s_1)}\pd{}{\s_4}\,,
\label{def_U_V_Z_x5_Routh_esf}	
\end{eqnarray}
and $\L_{12}\in C^\infty(\overline{M})$ is a non-vanishing function, 
are Poisson tensors of rank two in $\overline{M}$. 

The vector field $\overline{X}$ in $\overline{M}$, 
whose integral curves are the solutions of (\ref{sist_resf_sig4}),
is a Hamiltonian vector field with respect to the Poisson bivector
$\overline{\L}$ with the specific function $\L_{12}=(1-\s_1^2)/T(\s_1)$ and 
Hamiltonian function $\overline{E}$ given by (\ref{ener_resf_sig4}), 
i.e., $\overline{X}=\overline{\L}^\sharp(d\overline{E})$ in $\overline{M}$.
\label{prop_Lam_s_Routh_esf_s4}
\end{proposition}

\begin{proposition}
The bivectors $\L=f[(Z\phi)U\wedge V+Y\wedge Z]$, 
defined in ${\cal D}=\R^5$, where $U$ and $V$ are given 
as in Proposition~\ref{prop_Lam_s_Routh_esf_s4}, $Z=\partial{}/\partial{\s_5}$, 
$Y=(U\phi)V-(V\phi)U$, and $f\in C^\infty({\cal D})$ 
is a non-vanishing function, are Poisson tensors of rank two in ${\cal D}$,
except in the set of singular equilibria, where they vanish.

The vector field $X$ in ${\cal D}$, whose integral 
curves are the solutions of (\ref{sist_sig_5_routh_esf}),  
is a Hamiltonian vector field with respect to the Poisson bivector
$\L$ with the specific function $f=1/T(\s_1)$ and 
Hamiltonian function $E$ given by (\ref{ener_sig_5_routh_esf}), 
i.e., $X=\L^\sharp(dE)$ in ${\cal D}$.
\end{proposition}

Both Propositions can be proved as well by direct computations. 

\medskip
In this case, the equations (\ref{sys_chap_rs}) 
can be explicitly integrated in an easy way. 
{}From the second of these equations we have 
the relation $\s_4\sqrt{P(\s_1)}=k$.
Substituting into the first, we can also integrate to obtain 
the relation $I_1 r\s_3+I_3(\a+r \s_1)\s_4=j$. The 
constants $k,j$ are integration constants (essentially the 
initial conditions of the system (\ref{sys_chap_rs})). 
These two expressions are the desired first integrals 
(Casimir functions of the preceding Poisson structures).
The second of them is known as \emph{Jellet's integral}, 
see \cite{Cus98,EbeSch95} and references therein, see also
p. 184 of \cite{BorMam02c}. 

The invariant submanifolds thoroughly studied in \cite{Cus98}, 
could be interpreted in this framework as the symplectic leaves 
of the rank-two Poisson structure(s) $\overline{\L}$ of 
Proposition~\ref{prop_Lam_s_Routh_esf_s4}, determined by the
level sets of the first integrals $j$ and $k$.
 
\subsection{Ball rolling on a surface of revolution\label{ball_surf_revol}}

For this example we will follow the treatment and use some results
of \cite{Herm95b}, see therein for more details.
This problem has been treated as well, e.g., in 
\cite{BorMamKil02,NeiFuf72,Rou60}. In particular Routh, 
in the last of these references, noticed the existence of two integrals
of motion given by a system of two linear differential equations,
solved them in special cases, and described a family of stationary 
periodic motions together with a necessary condition for their stability.
Later, in \cite{Zen95}, it has been shown that the condition is also sufficient.
Both of \cite{Herm95b} and \cite{Zen95} prove that the corresponding
reduced system has integral curves consisting of either periodic orbits or 
equilibrium points.

Consider a homogeneous ball of mass $m$, radius $r$ and 
moment of inertia $M$ with respect to any principal axis. 
The ball rolls without slipping on a surface of revolution, 
under the influence of a vertical gravitational field of strenght $g$. 
We take the origin of coordinates at a point of the axis of symmetry 
of the surface (the intersection of this axis with the surface at its vertex), 
and we consider a horizontal plane passing through it. We
parametrize the position of the center of mass of the ball 
by its coordinates $(a_1,a_2)$ on this horizontal plane, 
and its height will be parametrized via
the smooth \emph{profile function} $\vp:\R\mapsto\R$ 
of the surface, $a_3=\vp\left(\sqrt{a_1^2+a_2^2}\right)$.
Note that not all surfaces of revolution can be parametrized well
in this way, e.g., the cylinder, which requires
a separate treatment, see Section~\ref{ball_in_a_cylinder} below. 
We will assume that $\vp$
is a smooth even function, thus we will have that
$\vp^{(2k+1)}(0)=0$, $k=0,\,1,\,2\,\dots$
We denote by $(\om_1,\om_2,\om_3)$ the components of the angular velocity 
vector $\om$ of the ball, and $(\g_1,\g_2,\g_3)$ the components of a unit 
vector $\g$ normal to the surface at the point of contact 
(directed towards the center of the ball). The unit vector
in the vertical coordinate axis is $e_3$.

The equations of motion can be easily computed by the classical equations
of the variation of the angular momentum, and implementing the non-holonomic
constraint of non-slipping of the point of contact, 
i.e., that its instantaneous velocity vanishes. 
They read (with respect to the center of mass of the ball, compare with 
eqs. (5), (7) of \cite{Herm95b} and Section~2 of \cite{BorMamKil02})
\begin{eqnarray}
&& M\frac{d\om}{dt}-m r^2\left(\frac{d}{dt}(\om\times\g)\right)\times\g-m g r e_3\times\g=0\,,
								\nonumber\\	
&& \dot a-r(\om\times\g)=0\,.					\label{eqs_rolling_ball}
\end{eqnarray}
The total mechanical energy of this system is then
\begin{equation} 
H=\frac12((M+m r^2)(\om\cdot\om)-m r^2(\g\cdot\om)^2)+m g a_3\,,
\label{ener_ball_surf}
\end{equation} 
and is a first integral for the system (\ref{eqs_rolling_ball}).

The system (\ref{eqs_rolling_ball}) and the energy (\ref{ener_ball_surf})
admit a further reduction of the $S^1$ symmetry consisting of 
rotations of the system about the vertical axis, 
and thus rotating both of $\om$ and $\g$ simultaneously. 
This action, as in the previous cases, is not free,
since the isotropy subgroup of pairs $((0,0,1),(0,0,\om_3))$ 
is $S^1$ (these pairs correspond to motions of the ball spinning 
around the vertical axis when being at the vertex of the surface), 
and we will use again invariant theory in order to perform the reduction, 
but now as it has been done in \cite{Herm95b}. 
First of all, we define the vector $v$ and the scalar $w$ as follows:
$v=r(\om\times\g)$, $w=-r(\om\cdot\g)$.
Then, a full set of invariant polynomials, which parametrize the orbit
space of the $S^1$ action, is 
\begin{eqnarray}
&& p_1=\frac12(a_1^2+a_2^2)\,,\quad p_2=a_1 v_1+a_2 v_2\,,
\quad p_3=a_1 v_2-a_2 v_1\,, 					\nonumber\\
&& p_4=w\,,\quad p_5=\frac12(v_1^2+v_2^2)\,, \label{inv_p}
\end{eqnarray} 
with the relations
\begin{equation}
p_2^2+p_3^2-4 p_1 p_5=0\,,\quad p_1\geq 0\,,\quad p_5\geq 0\,.
\label{rels_inv_pes}
\end{equation}
Therefore, the completely reduced orbit space $P$ is now 
the semialgebraic variety of $\R^5$ 
\begin{equation}
P=\{(p_1,\dots,p_5)\in\R^5\,\,|\,\,\phi(p)=0,\ p_1\geq 0,\ p_5\geq 0\}\,,
\end{equation}
where now $\phi\in C^\infty(\R^5)$ is the polynomial function 
$\phi(p)=p_2^2+p_3^2-4 p_1 p_5$. $P$ is \emph{not} a smooth submanifold 
of $\R^5$, because the previous $S^1$ action is not free. Instead,
$P$ is homeomorphic to a cone in $\R^4$ times $\R$ \cite{Herm95b},
which can be easily seen from the relation $\phi(p)=0$ when it is 
written as $p_2^2+p_3^2+(p_1-p_5)^2=(p_1+p_5)^2$. The vertex of the cone
is determined by $p_2=p_3=p_1=p_5=0$, therefore 
the singular points of $P$ are 
$$
\Pi=\{(0,0,0,p_4,0)\in P\,\,|\,\,p_4\in\R\}\,.
$$

Calculating the time derivatives of the invariants, 
using (\ref{eqs_rolling_ball}), and the relations
$$
\g_1=-\frac{a_1}{\sqrt{2 p_1}}\frac{\vp'}{\sqrt{1+\vp'^2}}\,,\quad
\g_2=-\frac{a_2}{\sqrt{2 p_1}}\frac{\vp'}{\sqrt{1+\vp'^2}}\,,\quad
\g_3=\frac{1}{\sqrt{1+\vp'^2}}\,,
$$
(we will use the notation $\vp=\vp(\sqrt{2 p_1})$, $\vp'=\vp'(\sqrt{2 p_1})$ 
and $\vp''=\vp''(\sqrt{2 p_1})$ in what follows) we arrive to the system in the
reduced orbit space $P$
\begin{eqnarray}
\fl \quad\quad \dot p_1=p_2						\nonumber\\
\fl \quad\quad \dot p_2=\frac{1}{1+\vp'}
\left\{
-\frac{M}{\a r^2} p_3 p_4 \frac{\vp'}{\raiz}
-\frac{m g}{\a}\raiz\vp'+2 p_5
-p_2^2\frac{\vp'}{\raiz}\left(\vp''-\frac{\vp'}{\raiz}\right)
\right\}						\nonumber\\
\fl \quad\quad \dot p_3=\frac{M}{\a r^2} p_2 p_4 \frac{\vp''}{1+\vp'^2}
							\label{eqs_Herm}\\
\fl \quad\quad \dot p_4=-\frac{p_2 p_3}{2 p_1}\left(\frac{\vp''}{1+\vp'^2}-\frac{\vp'}\raiz\right) 
							\nonumber\\
\fl \quad\quad \dot p_5=\frac{p_2}{1+\vp'}
\left\{
\frac{1}{2 p_1}\left(\frac{M}{\a r^2} p_3 p_4-p_2^2\frac{\vp'}{\raiz}\right)
\left(\vp''-\frac{\vp'}{\raiz}\right)
-\frac{m g}{\a}\frac{\vp'}{\raiz}-2 p_5 \frac{\vp'^2}{2 p_1}
\right\}							\nonumber
\end{eqnarray}
and the reduced energy  
\begin{equation}
E=\frac{M}{2 r^2} p_4^2+\a p_5+\frac{\a\vp'^2}{4 p_1} p_2^2+m g \vp\,,
\label{Ered_Herm}
\end{equation}
where $\a=\frac{M+mr^2}{r^2}$. These are the equations found in 
Lemmas 2.2 and 2.3 (i) of \cite{Herm95b}.

We observe that the right hand sides of (\ref{eqs_Herm}) 
and (\ref{Ered_Herm}) make sense in an open set ${\cal D}$ of $\R^5$ 
larger than $P$, namely ${\cal D}=\R^5\backslash\{(p_1,p_2,p_3,p_4,p_5)\ |\ p_1<0\}$. 
This is due to the fact that they are defined in the limit $p_1\rightarrow 0^+$,
because of the assumption that the odd-order derivatives at 0 of $\vp$ vanish. 
(For points strictly in $P$ with $p_1=0$ this assumption would not be 
necessary, since these points also have $p_2=0$, $p_3=0$).
We will consider the enlarged domain ${\cal D}$ for the vector field $X$
whose integral curves are the solutions of (\ref{eqs_Herm}), and also for 
the reduced energy (\ref{Ered_Herm}), compare with p. 500 of \cite{Herm95b}.

The regular stratum of $P$, i.e., $P\backslash\Pi$, can be covered 
by two charts \cite{FasGiaSan04}, whose corresponding neighbourhoods 
can be chosen to be the smooth open dense subsets $\overline{P}_1,\overline{P}_2\subset P$ 
given by
\begin{eqnarray}
&& \overline{P}_1
=\left\{
(p_1,p_2,p_3,p_4,p_5)\in\R^5\,\,\bigg|\,\,p_5=\frac{p_2^2+p_3^2}{4 p_1},\ p_1>0
\right\}\,,
\label{smooth_subman_P} \\
&& \overline{P}_2
=\left\{
(p_1,p_2,p_3,p_4,p_5)\in\R^5\,\,\bigg|\,\,p_1=\frac{p_2^2+p_3^2}{4 p_5},\ p_5>0
\right\}\,. \nonumber
\end{eqnarray}

However, for our purposes here, it will be enough to consider 
just $\overline{P}_1$, in order to endow it with Poisson structures of the type 
described in Theorem~\ref{propo_Pois_x4}, which afterwards could be compared with
the Poisson structure given originally in \cite{BorMamKil02}. The 
procedure for $\overline{P}_2$ is analogous. Thus, we will consider
the induced vector field $\overline{X}$ and energy $\overline{E}$ 
on $\overline{P}_1$, which can be found
{}from (\ref{eqs_Herm}) and (\ref{Ered_Herm}) by solving for $p_5$. 
The integral curves of $\overline{X}$ are the solutions of the system
\begin{eqnarray}
\fl \quad\dot p_1=p_2						\nonumber\\
\fl \quad\dot p_2=\frac{1}{1+\vp'}
\bigg\{
-\frac{M}{\a r^2} p_3 p_4 \frac{\vp'}{\raiz}
-\frac{m g}{\a}\raiz\vp' 				
+\frac{p_2^2+p_3^2}{2 p_1}
-p_2^2\frac{\vp'}{\raiz}\left(\vp''-\frac{\vp'}{\raiz}\right)
\bigg\}						\nonumber\\
\fl \quad\dot p_3=\frac{M}{\a r^2} p_2 p_4 \frac{\vp''}{1+\vp'^2}
							\label{eqs_Herm_rd_p4}\\
\fl \quad\dot p_4=-\frac{p_2 p_3}{2 p_1}\left(\frac{\vp''}{1+\vp'^2}-\frac{\vp'}\raiz\right) 
							\nonumber
\end{eqnarray}
meanwhile 
\begin{equation}
\overline{E}=\frac{M}{2 r^2} p_4^2+\a \frac{p_2^2+p_3^2}{4 p_1}
+\frac{\a\vp'^2}{4 p_1} p_2^2+m g \vp\,.
\label{Ered_Herm_rd_p4}
\end{equation}

Now, the reduced vector field $X$ satisfies $X(E)=0$ and $X(\phi)=0$ 
in ${\cal D}$, and $\overline{X}(\overline{E})=0$ in $\overline{P}_1$.
The vector field $X$ has a family of singular equilibrium points
consisting of the singular set $\Pi$, that is, $\{(0,0,0,p_4,0)\ |\ p_4\in\R\}$, 
which as already mentioned, correspond to the spinning of the ball 
about the vertical when being at the vertex of the surface
(then the reduced energy becomes $\frac{M}{2 r^2}p_4^2+mg\vp(0)$).
$X$ has as well a family of regular equilibria given by the set of constants 
$$
\left\{(p_{10},0,p_{30},p_{40},p_{50})\in {{\cal D}}\ \bigg|\ 
2 p_{50}-\frac{mg}{\a}\sqrt{2 p_{10}}\vp'(\sqrt{2 p_{10}})-\frac{M}{\a r^2}
\frac{\vp'(\sqrt{2 p_{10}})}{\sqrt{2 p_{10}}}p_{30}p_{40}=0
\right\}\,.
$$ 
These regular equilibria correspond in the original system to rotations
of the ball along a parallel of the surface of revolution at constant height. 

In addition, both of the systems (\ref{eqs_Herm}) and (\ref{eqs_Herm_rd_p4})
admit two first integrals of motion related to the solutions (in the sense explained 
in Section~\ref{Pois_struc_R4}) of the non-autonomous 
linear system (see also \cite{Herm95b}, Lemma 2.3. (ii))
\begin{eqnarray}
\fl\quad\quad\quad
\frac{dp_3}{dp_1}=\frac{M}{\a r^2} p_4 \frac{\vp''}{1+\vp'^2}\,,
\quad
\frac{dp_4}{dp_1}=-\frac{p_3}{2 p_1}\left(\frac{\vp''}{1+\vp'^2}-\frac{\vp'}\raiz\right)\,, 
\label{Chap_mat}
\end{eqnarray}
Let $\th_0=d\phi$ and $\th_1$, $\th_2$, be the 1-forms, defined in $\overline{P}_1$
(resp. ${\cal D}$) by 
\begin{eqnarray}
\fl\quad\quad\quad
\th_1=\frac{M}{\a r^2} p_4 \frac{\vp''}{1+\vp'^2} dp_1-dp_3\,,
\quad
\th_2=\frac{p_3}{2 p_1}\left(\frac{\vp''}{1+\vp'^2}-\frac{\vp'}\raiz\right) dp_1+dp_4\,.
							\nonumber
\end{eqnarray} 
We have the following results, applying the Theorems of Section~\ref{Pois_struc_R4_R5},
which can be proved by direct computations:
\begin{proposition}
The bivectors of the form $\overline{\L}=-\L_{12}\,U\wedge V$, 
defined in $\overline{P}_1$, where 
\begin{eqnarray}
\fl\quad\quad U=\pd{}{p_2}\,,\quad V=\pd{}{p_1}
+\frac{M}{\a r^2} p_4\frac{\vp''}{1+\vp'^2}\pd{}{p_3}
-\frac{p_3}{2 p_1}\left(\frac{\vp''}{1+\vp'^2}-\frac{\vp'}\raiz\right)\pd{}{p_4}\,,
\label{def_U_V_Z_x5_ball_surf}
\end{eqnarray}
and $\L_{12}\in C^\infty(\overline{P}_1)$ is a non-vanishing function, 
are Poisson tensors of rank two in $\overline{P}_1$. 

The vector field $\overline{X}$ in $\overline{P}_1$, 
whose integral curves are the solutions of (\ref{eqs_Herm_rd_p4}),
is a Hamiltonian vector field with respect to the Poisson bivector
$\overline{\L}$ with the specific function $\L_{12}={2 p_1}/{\a(1+\vp'^2)}$ 
and Hamiltonian function $\overline{E}$ given by (\ref{Ered_Herm_rd_p4}), 
i.e., $\overline{X}=\overline{\L}^\sharp(d\overline{E})$ in $\overline{P}_1$.
\label{Hamilton_eqs_roll_ball_surf_p4}
\end{proposition}

\begin{proposition}
The bivectors $\L=f[(Z\phi)U\wedge V+Y\wedge Z]$, 
defined in ${\cal D}\subset\R^5$, where $U$ and $V$ are given 
by (\ref{def_U_V_Z_x5_ball_surf}), $Z=\partial{}/\partial{p_5}$, 
$Y=(U\phi)V-(V\phi)U$, and $f\in C^\infty({\cal D})$ 
is a non-vanishing function, are Poisson tensors of rank two in ${\cal D}$,
except in the set of singular equilibria, where they vanish.

The vector field $X$ in ${\cal D}$, whose integral 
curves are the solutions of (\ref{eqs_Herm}),  
is a Hamiltonian vector field with respect to the Poisson bivector
$\L$ with the specific function $f={1}/{2\a(1+\vp'^2)}$ and 
Hamiltonian function $E$ given by (\ref{Ered_Herm}), 
i.e., $X=\L^\sharp(dE)$ in ${\cal D}$.
\end{proposition}

\medskip 
{\noindent {\bf Remarks}} For the present case, 
a Poisson structure analogous to one of the structures $\overline{\L}$ of 
Proposition~\ref{Hamilton_eqs_roll_ball_surf_p4} has been found, 
to the best of our knowledge, by the first time in \cite{BorMamKil02}, 
see their equation (3.11) for $\l=0$. 
In fact, up to a rescaling, they are the same, by using the identifications
\begin{eqnarray}
\fl\quad\quad 
x_2=\frac M r p_4\sqrt{2 p_1}\frac{\sqrt{1+\vp'^2}}{\vp'}\,,\quad 
x_3=-\frac M r \frac{\sqrt{2 p_1} p_4+p_3\vp'}{\sqrt{2 p_1}\sqrt{1+\vp'^2}}\,,
\quad
x_4=\a r\frac{p_2\sqrt{1+\vp'^2}}{\sqrt{2 p_1}}\,,\nonumber\\
\fl\quad\quad 
x_1=\frac{1}{\sqrt{1+\vp'^2}}\,,\quad		
f(x_1)=-\frac{\sqrt{2 p_1}\sqrt{1+\vp'^2}}{\vp'}\,.	\nonumber
\end{eqnarray}
The (local) Poisson bivector found in \cite{BorMamKil02} for this case 
reads in their coordinates $(x_1,x_2,x_3,x_4)$ as
$$
\left\{
\a r^2
\left(\pd{}{x_1}+\frac{f'(x_1)}{x_1}x_3\pd{}{x_2}\right)
+m r^2\frac{x_2}{f(x_1)}\pd{}{x_3}
\right\}\wedge\pd{}{x_4}\,,
$$
which, in particular, is also of the type described in Example~1. 
Therefore, multiples of this bivector are again Poisson bivectors
and hence, the rescaling introduced in \cite{BorMamKil02}, by means of
an \emph{invariant measure}, in order to render the reduced system 
Hamiltonian, is unnecessary.

On the other hand, Hermans in \cite{Herm95b} has not noticed 
the existence of any of these Poisson structures of rank two 
but he constructed a closed 2-form, with domain contained
in $\overline{P}_1$, which vanish in a set containing 
the set of regular equilibria, but has rank four otherwise.
For this construction, which uses \emph{non-holonomic reduction} 
\cite{BatSni92b}, it is indeed \emph{necessary to rescale} 
the original reduced vector field, see Section~4.1 of \cite{Herm95b}.

\subsection{Ball rolling on the interior of a cylinder\label{ball_in_a_cylinder}}

In this Section we will treat the special case of a ball 
rolling inside of a cylinder, which cannot be parametrized 
as in Section~\ref{ball_surf_revol}. 
In contrast with the general case,
this case is completely and explicitly solvable, as it is well known,  
see, e.g., \cite{NeiFuf72,BatGraMac96,BorMamKil02,Marle03}.
However, we will give an independent treatment.

For this specific system, we will easily find a family of Poisson structures 
of rank two, generated by two of them, with respect to which the reduced system 
is Hamiltonian with the reduced energy as Hamiltonian function. 

Consider therefore the case of the ball rolling inside a surface of 
revolution, with the following variations: the center of mass of
the ball will be parametrized by the vector $a$, with cylindrical 
coordinates $(\r \cos\th,\r \sin\th,z)$, where $\r$ is the radius
of the cylinder on which the center of mass of the ball moves, and
$z$ is the height with respect to the gravitational energy reference point.
The normal vector $\g$ reads then
$\g=-(\cos\th,\sin\th,0)$.
The system (\ref{eqs_rolling_ball}) becomes in the 
coordinates $(\om_1,\om_2,\om_3)$ and $(\th,z)$ 
\begin{eqnarray}
&& \dot\om_1=\frac m\a\left(\frac g r
+\frac r \r\om_3(\om_1\cos\th+\om_2\sin\th)\right)\sin\th\,, 	\nonumber\\
&& \dot\om_2=-\frac m\a\left(\frac g r
+\frac r \r\om_3(\om_1\cos\th+\om_2\sin\th)\right)\cos\th\,, 	\label{syst_cyl}\\
&& \dot\om_3=0\,,
\quad\dot\th=-\frac r\r\om_3\,,
\quad\dot z=r(\om_2\cos\th-\om_1\sin\th)\,.			\nonumber
\end{eqnarray} 
Likewise, the energy (\ref{ener_ball_surf}) reads now
\begin{equation} 
H=\frac12\{(M+m r^2)(\om\cdot\om)-m r^2(\om_1\cos\th+\om_2\sin\th)^2\}+m g z\,,
\label{ener_ball_surf_cyl}
\end{equation} 
which is conserved by the system (\ref{syst_cyl}). Obviously, $\om_3$ is a 
first integral of the system as well. 

Let us consider now the system obtained after the reduction of the $S^1$ 
symmetry of rotations of the whole system about the vertical axis, as in the
general case. Although now the $S^1$ action is free, we will 
use again invariant theory in order to perform the reduction.
Consider the invariants similar (but not equal) to (\ref{inv_sigma}):
\begin{eqnarray}
&& \s_1=z\,,
\quad\s_2=\g_1\om_2-\g_2\om_1=-\om_2\cos\th+\om_1\sin\th\,, \nonumber\\
&& \s_3=\g_1\om_1+\g_2\om_2=-\om_1\cos\th-\om_2\sin\th\,, \quad\s_4=\om_3\,,
\label{inv_sigma_ball_cyl}
\end{eqnarray}
which in this case can be regarded as coordinates on $\R^4$. 
Then, the reduced system for $(\s_1,\s_2,\s_3,\s_4)$ reads
\begin{eqnarray}
\dot\s_1=-r \s_2\,,
\quad\dot\s_2=\frac{M \s_4}{\a r\r}\s_3+\frac{m g}{\a r}\,,
\quad\dot\s_3=-\frac r\r\s_4\s_2\,,
\quad\dot\s_4=0\,,
\label{sist_red_sigma_ball_cyl}
\end{eqnarray} 
which preserves the reduced energy
\begin{equation}
E=\frac12\{mr^2(\s_2^2+\s_4^2)+M(\s_2^2+\s_3^2+\s_4^2)\}+mg\s_1\,.
\label{E_red_ball_cyl}
\end{equation}
The reduced vector field $X$ in the reduced space reads then
\begin{equation}
X=-r \s_2\pd{}{\s_1}
+\left(\frac{M \s_4}{\a r\r}\s_3+\frac{m g}{\a r}\right)\pd{}{\s_2}
-\frac r\r\s_4\s_2\pd{}{\s_3}\,,
\label{red_vf_ball_cyl}
\end{equation}
and we have $X(E)=0$ in all points of $\R^4$.
The general solution of (\ref{sist_red_sigma_ball_cyl}) can be given explicitly. 
It reads
\begin{eqnarray}
&& \fl \quad\quad \s_1(t)=\s_1(0)-\frac r{\n_1\n_2}\{\s_2'(0)(1-\cos \sqrt{\n_1\n_2}t)
+\sqrt{\n_1\n_2}\,\s_2(0)\sin\sqrt{\n_1\n_2} t\}
								\nonumber\\
&& \fl \quad\quad \s_2(t)=\s_2(0)\cos\sqrt{\n_1\n_2} t
+\frac{\s_2'(0)}{\sqrt{\n_1\n_2}}\sin\sqrt{\n_1\n_2} t 		
						\label{sol_sist_red_sigma_ball_cyl}\\
&& \fl \quad\quad \s_3(t)=-\frac{\s_g}{\n_2}+\frac{\s_2'(0)}{\n_2}\cos\sqrt{\n_1\n_2} t
-\sqrt{\frac{\n_1}{\n_2}}\,\s_2(0)\sin\sqrt{\n_1\n_2} t		\nonumber\\
&& \fl \quad\quad \s_4(t)=\s_4\,,						\nonumber
\end{eqnarray} 
where we have defined the constants $\n_1=r \s_4/\r$, 
$\n_2=M \n_1/{\a r^2}$ 
and
$\s_g={m g}/{\a r}$. 
It is clear that the reduced system, if $\s_4\neq0$, has integral curves consisting of 
either periodic orbits or equilibrium points, belonging to the set
$\{(\s_{10},0,-m g\r/M\s_{40},\s_{40})\in\R^4\ |\ \s_{10}, \s_{40}\in\R,\ \s_{40}\neq0\}$.
These equilibrium points correspond to rotations of the ball inside the cylinder
at constant height, as in the general case. 
On this occasion, the reduced system can be reconstructed easily 
to the complete system, thus the general solution of (\ref{syst_cyl}) is
\begin{eqnarray}
&&\om_1(t)=\s_2(t)\sin\th(t) -\s_3(t)\cos\th(t)\,,  \nonumber\\
&&\om_2(t)=-\s_2(t)\cos\th(t) -\s_3(t)\sin\th(t)\,, \label{sol_sist_sigma_ball_cyl}\\
&&\om_3(t)=\s_4=\om_3(0)\,,\quad z(t)=\s_1(t)\,,	\nonumber
\end{eqnarray} 
where $\th(t)=\th_0-\n_1 t$ and $\s_i(t)$, $i=1,2,3$ are given by 
(\ref{sol_sist_red_sigma_ball_cyl}). If we denote $\om_1(0)=\om_{10}$, 
$\om_2(0)=\om_{20}$, we have the relations for the initial conditions
\begin{eqnarray}
\fl \quad\quad \s_2(0)=\om_{10}\sin\th_0-\om_{20}\cos\th_0\,,
\quad\s_2'(0)=\s_g-\n_2(\om_{10}\cos\th_0+\om_{20}\sin\th_0)\,.	\nonumber
\end{eqnarray} 
The complete system, when $\om_3\neq0$,
is then isochronus with two frequencies, the motions being periodic 
(relative equilibria, projecting to equilibrium points in the reduced space) 
or quasi-periodic, otherwise. The solutions with $\om_3=0$ correspond
to \emph{falling motions} of the ball, rolling along a vertical 
generatrix of the cylinder. The explicit expression of these solutions is
\begin{eqnarray}
&&\om_1(t)=\om_{10}+t\,\s_g\sin\th_0\,,
\quad\om_2(t)=\om_{20}-t\,\s_g\cos\th_0\,,
\quad\om_3(t)=0\,,				\nonumber\\
&&\th(t)=\th_0\,,
\quad z(t)=z_0-\frac12r\s_gt^2+rt\,(\om_{20}\cos\th_0-\om_{10}\sin\th_0)\,.
						\nonumber
\end{eqnarray} 

We will treat now the question of writting the vector field $X$, 
given by (\ref{red_vf_ball_cyl}), as a Hamiltonian vector field with
respect to a Poisson structure of rank two, with Hamiltonian function $E$.

We first observe that the reduced vector field $X$ is annihilated by the 1-forms
\begin{eqnarray}
\fl\quad\quad \th_1=d\s_4\,,\quad\th_2=-\frac{\s_4}{\r}d\s_1+d\s_3\,,
\quad\th_3={\a r^2\s_2\s_4}d\s_2+({M\s_3\s_4+m g\r})d\s_3\,,\nonumber
\end{eqnarray}
and then, it is easy to apply Theorem~\ref{propo_Pois_x4}, 
to obtain the following results:
\begin{proposition} 
The bivector 
$\L_1=\pd{}{\s_2}\wedge\frac1{\s_2}X
=-\pd{}{\s_2}\wedge\left(r\pd{}{\s_1}+\frac{r\s_4}\r\pd{}{\s_3}\right)$ is a Poisson 
bivector on $\R^4$ of rank two such that $\L_1^\sharp(\th_1)=\L_1^\sharp(\th_2)=0$.
Likewise, The bivector 
$\L_2=-\frac{1}{mg}\pd{}{\s_1}\wedge X
=-\frac{1}{mg}\pd{}{\s_1}\wedge\left[
\left(\frac{M \s_4}{\a r\r}\s_3+\frac{m g}{\a r}\right)\pd{}{\s_2}
-\frac r\r\s_4\s_2\pd{}{\s_3}\right]$ 
is a Poisson bivector on $\R^4$ of rank two such that 
$\L_2^\sharp(\th_1)=\L_2^\sharp(\th_3)=0$. In addition, we have 
$X=\L_1^\sharp(dE)=\L_2^\sharp(dE)$, where $X$ is given 
by (\ref{red_vf_ball_cyl}) and $E$ by (\ref{E_red_ball_cyl}).
\label{prop_Pois_1_cyl}
\end{proposition}

Now, the Pfaff systems \lq\lq $\th_1=0,\th_2=0$\rq\rq\ and \lq\lq $\th_1=0,\th_3=0$\rq\rq\  
can be easily integrated, giving non-trivial Casimir functions 
of $\L_1,\L_2$, and first integrals of $X$:

\begin{proposition} We have 
${\rm ker}\,\L_1^\sharp={\rm span}\{dc_1,dc_2\}$, and 
${\rm ker}\,\L_2^\sharp={\rm span}\{dc_1,dc_3\}$, 
where $c_1=\s_4$, $c_2=\s_3-\frac{\s_4}{\r}\s_1$ 
and $c_3=\frac{r \s_4}{\r}\s_2^2+
\left(\frac{M \s_4}{\a r\r}\s_3+\frac{m g}{\a r}\right)\s_3$. 
\label{prop_Cas_Pois_1_cyl}
\end{proposition}

As a consequence, we have that the reduced vector field $X$ has in principle
four first integrals, namely, $E$, $c_1$, $c_2$ and $c_3$, but clearly, 
they form a functionally dependent set. However, for example, we have that 
$\{E,c_2,c_3\}$ is generically an independent set of first integrals, 
although in the equilibrium points one becomes dependent of the other two. 
In the falling motions, $\s_4=0$, therefore $\L_1$ and $\L_2$ become proportional.  

Incidentally, we also observe that $\L_1^\sharp(dc_3)=\frac1{\a r^2}X$ and 
$\L_2^\sharp(dc_2)=-\frac{\s_4}{\r mg}X$. In addition, the Poisson 
bivectors $\L_1$, $\L_2$ are compatible in the sense that their
Schouten--Nijenhuis bracket vanishes, $[\L_1,\L_2]=0$, which can be 
checked, e.g., using the properties (\ref{props_Scho}). Thus we have
the following result:
   
\begin{proposition} 
The pencil of bivectors $\L_\l=(1-\l)\L_1+\l\L_2$ consists of Poisson bivectors
of rank two such that $X=\L_\l^\sharp(dE)$ for all $\l\in\R$. Moreover, 
the functions $c_1$, $c_{2\l}=(1-\l)E-\a r^2 c_3$ and 
$c_{3\l}=\l\,{r\s_4}E/\r+mg r\,c_2$ are (functionally dependent) 
Casimir functions of $\L_\l$ (and therefore, first integrals of $X$) for all $\l\in\R$. 
\label{prop_Cas_Pois_lambda_cyl}
\end{proposition}

{\noindent {\it Proof}} 

That the rank of $\L_\l$, for all $\l\in\R$, is two, 
is obvious when one realizes that it does not contain terms 
on $\pd{}{\s_4}$ and therefore the rank must be an even number
between zero and four. 
The other statements are a matter of computation using the
above observations.
 
\section{Conclusions and outlook}

We have shown the form of certain Poisson structures of rank two with
respect to which certain reduced problems of non-holonomic mechanics
become Hamiltonian. We have shown that in $\R^4$ and $\R^5$, 
from an algebraic point of view, these Poisson structures are defined,
up to a factor function, by the choice of the kernel of bivectors on these spaces
to be generated by 1-forms of a specific type. Such 1-forms
define integrable codistributions in the sense of Frobenius,
and are chosen in order to accommodate and generalize
the systems of first order non-autonomous differential equations
which appear after reduction in certain non-holonomic mechanical systems, 
whose solutions are related to first integrals of such reduced systems.

We have applied the theory to the cases of the rolling disk, 
the Routh's sphere, and the ball rolling on a surface of revolution, 
explicitly recovering as a particular case some results of \cite{BorMamKil02}. 
Thus, we have shown that the framework suggested by 
Borisov, Mamaev and Kilin \cite{BorMam02c,BorMamKil02} can 
be improved along the lines discussed, namely, that those
reduced systems need no rescaling to become Hamiltonian
with respect to a Poisson structure of rank two, and
that the domain of definition of the Poisson structures 
introduced therein can be extended, including even the set of 
singular equilibria of the reduced systems.
A natural question is whether a similar approach could be used in 
other non-holonomic systems, maybe of higher dimension.

However, there are more fundamental points still to be 
better understood. For example, to what extent the mentioned 
Poisson structures can be useful to investigate the intimate nature 
of these and maybe other non-holonomic systems, for example in order 
to characterize their integrability properties \cite{BatCus99,Fas98,FasGia02}; 
see also the recent work \cite{FasGiaSan04}.
Another question could be to clarify the origin of the system 
of differential equations giving the conservation laws for the
mentioned reduced non-holonomic systems, 
see also \cite{BatGraMac96,CanLeoMarrMar98,CusKemSniBat95,Marle03,Sni02} and 
references therein.

\section*{Acknowledgements}

This work is part of the research contract HPRN-CT-2000-00113, 
supported by the European Commission funding for the Human Potential
Research Network \lq\lq Mechanics and Symmetry in Europe\rq\rq\ (MASIE). 

The author is specially indebted to F. Fass\`o for valuable 
comments, questions and ideas, also concerning previous versions of this article. 
Likewise, the author acknowledges useful comments
and warm hospitality from J.F. Cari\~nena, R. Cushman and T. Ratiu 
at their respective institutions 
(Universidad de Zaragoza, University of Utrecht and EPFL).

\end{document}